%%%% 
\input amstex
%\input epsf
%\input preamble
%\newsymbol\boxtimes 1202

%letters

\def\b1{\text{\bf 1}}

\def\dpar{\partial}

\def\#{\,\check{}}

%symbols

% space between paragraphs 
%\parskip = 6pt
\predefine\divs{\div}

%operators
\def\rot{\operatorname{rot}}
\def\div{\operatorname{div}}
\def\pderby#1{{\dpar}\over{\dpar{#1}}}
\def\psderby#1{{\dpar^2}\over{\dpar{#1}^2}}

\documentstyle{amsppt}

% \LimitsOnInts
% \LimitsOnLimits

\NoBlackBoxes

\topmatter
\title Quantum mechanics of the free Dirac electrons and Einstein photons, and the Cauchy process \endtitle
\author A.~A.~Beilinson  \endauthor
\leftheadtext{A.~A.~Beilinson}

\address 
\endaddress

\email  \endemail

\thanks \endthanks

%  Math Subject Classifications 
%\subjclassyear{2}
%\subjclass  \endsubjclass


\abstract Fundamental solutions for the free Dirac electron and Einstein photon equations  in position coordinates are constructed as matrix valued functionals on the space of bump functions. It is shown that these fundamental solutions are related by a unitary transform via the Cauchy distribution in imaginary time. We study the way the classical relativistic mechanics of the free particle comes from the quantum mechanics of the free Dirac electron.   \endabstract

\endtopmatter

\document
%\magnification = 1100

%\nologo

\head  Introduction \endhead

We study fundamental solutions for the Dirac electron equations and the Maxwell equations for electromagnetic field without sources (the photon field) as functionals of space variables, the time is viewed as a parameter.
\medskip

%or neutrino  including the   case of $m = 0$, and their interrelations. Here $\gamma^\mu = \{\gamma^0 ,\gamma \}$ are Dirac matrices (see e.g.~\cite{8}), $m$ is the mass of the Dirac particle; $t\ge 0$ is the physical time, $x = \{ x_1 , x_2 , x_3 \} $ are position coordinates.

%We also study fundamental solutions for the Maxwell equations for electromagnetic field without sources (see \cite{1})  where $E_t (x)$ and $H_t(x)$ are electric and magnetic fields, or the field of photon.

We first construct the fundamental solutions as analytic functionals in position coordinates, so in momentum coordinates these are functionals on bump functions; then, using a renormalization procedure, we construct the position coordinates presentation of our fundamental solutions as functionals on bump functions.
\medskip

Specifically, we exploit the Foldy-Wouthuysen presentation of solutions (diagonalized fundamental solutions) of the Dirac and Maxwell equations that reduces them to a scalar transition probability of the Cauchy process in imaginary time, using both momentum and position coordinate presentations of this generalized function.
\medskip

For a short exposition of main results see \cite{15}.

\head 1.  A particular case of the Dirac equation with $m = 0$.  \endhead

Here  the Dirac equation looks as (see e.g.~\cite{8})
$$
\gamma^0 {\pderby t} + (\gamma,\nabla) = 0. 
\tag 1
$$
We use the system of units where the Planck constant $\hbar$ and the velocity of light $c$ are equal to 1. Denote by $D_t(x)$ the fundamental solution of (1);  its momentum coordinate presentation $\tilde{D}_t(p)$ is then
$$
\tilde{D}_t(p) = \exp(it\gamma^0 (\gamma,p)).
\tag 2
$$
Notice that the matrix $\gamma^0 (\gamma,p)$ in (2) is Hermitian, hence it can be diagonalized; it is easy to see that this can be done using a unitary (and Hermitian) operator $\tilde{T}(p) = \tilde{T}^{-1}(p) = {{\gamma^0}\over{\sqrt{2}}}((\gamma,p_e) + I)$ where $p_e$ is the unit norm vector for $p$, thus $\gamma^0 (\gamma,p) = \tilde{T}(p) \gamma^0 \rho \tilde{T}(p)$ and
$$
\tilde{D}_t(p) = \tilde{T}(p)\exp(it\gamma^0 \rho)\tilde{T}(p),
\tag 3
$$
where $\rho = \sqrt{p_1^2 + p_2^2 + p_3^2}$.
\medskip

Therefore (3) provides  the fundamental solution of the Dirac equation (1) in momentum space in the Foldy-Wouthuysen variables (see \cite{13}) 
$$
\tilde{D}^F_t(p) = \exp(it\gamma^0 \rho). 
\tag 4
$$
Notice that the Foldy-Wouthuysen transform of solutions of the Dirac equation   is an isomorphism.
\medskip

The operator in (4) is diagonal, hence it reduces to two scalar unitary operators that are complex conjugate 
$$
\tilde{C}_{it}(p) = \exp(it\rho), \quad \overline{\tilde{C}}_{it}(p) = \tilde{C}_{-it}(p) = \exp(-it\rho). 
\tag 5
$$
Our problem is to write down the position coordinate presentation of  these operators.
\medskip

Recall that a probability density $1\over{\pi(1 + x_1^2)}$, whose Fourier transform is $\exp(-|p_1|)$, was first studied by Cauchy; the corresponding 1-dimensional process has  transition probability ${t\over{\pi (t^2 + x_1^2 )}} = C_t(x_1)$, $t\ge 0$;  the Cauchy process in 3-dimensional space has   transition probability 
$$
{t\over{\pi^2  (t^2 + r^2 )^2}} = C_t (x), 
\tag 6
$$
where $r = \sqrt{x_1^2 + x_2^2 + x_3^2}$, $t\ge 0$, with  the momentum presentation $\exp(-t\rho)$, $\rho = \sqrt{p_1^2 + p_2^2 + p_3^2}$ (see \cite{7}). Notice that $C_t(x)\neq \mathop\prod\limits_{j = 1}^{3} C_t(x_j)$ and that process is not Gaussian. So the passage from the 3-dimensional problem $C_{it}(x)$ to the 1-dimensional one $C_{it}(x_1)$ can be performed only through integration $C_{it}(x_1) = \int C_{it}(x)dx_2 dx_3$.
\medskip

Therefore the construction of the position space presentation of operators (5) is reduced to a correct analytic continuation of operator (6) from real positive time to the imaginary one, which is possible if $C_{it}(x)$ and $\tilde{C}_{it}(p)$ are understood as generalized functions (see \cite{2}).

\head  2. One-dimensional Cauchy distribution in imaginary time \endhead

Consider first one-dimensional case with the momentum space retarded Green's function $\exp(it|p_1|) = \tilde{C}_{it}(p)$ which is the analytic continuation to imaginary time of the momentum space transition probability of the Cauchy process $\tilde{C}_{t}(p) = \exp(-t|p_1|)$, $t\ge 0$. Therefore the result is not Gaussian as well; it is natural to call it one-dimensional Cauchy process in imaginary time.
\medskip

We assume first that the functional $\tilde{C}_{it}(p)$ is defined on the space of bump test functions $\varphi(p) \in K^{(1)}$, and the retarded Green's function $C_{it}(x)$ is analytic functional on $Z^{(1)}$ (see \cite{2}).
\medskip

Consider  Parseval's identity 
$$
\int \overline{C_{it}(x)}\psi(x)dx = {1\over{2\pi}}\int \overline{\exp(it|p|)}\varphi(p)dp. 
\tag 7
$$
where $\varphi(p) \in K^{(1)}$ is a bump function and $\psi(x) \in Z^{(1)}$ is its Fourier transform which is an entire function of order 1 (Paley-Wiener theorem, see \cite{2}).
\medskip

Thus the r.h.s.~in (7) equals 
$$
{1\over{2\pi}}\int \exp(-it|p| + ixp)\psi(x)dpdx. 
\tag 8
$$

Since $\exp(it|p|)$ and $\psi(x)$ are entire functions and $\psi(x)$ decays rapidly at infinity near the real axis (see \cite{2}), the integral along real axis $x$ in (8) equals the integral along any axis parallel to the real one. It is easy to see that one can choose the integration path so that the integral by $p$ converges absolutely (and its value can be found in a list of integrals). Indeed, (8) becomes  

$$
\multline
\left.{1\over{2\pi}}\int \int\limits_{-\infty}^0 e^{(itp + ipx)}dp\psi(x)dx\right|_{x \to x - i0} + \left.{1\over{2\pi}}\int \int\limits_0^{\infty} e^{(-itp + ipx)}dp\psi(x)dx\right|_{x \to x + i0} = \\ 
= \left.{1\over{2\pi}}\int \int\limits_0^{\infty} e^{(-itp - ipx)}dp\psi(x)dx\right|_{x \to x - i0} + \left.{1\over{2\pi}}\int \int\limits_0^{\infty} e^{(-itp + ipx)}dp\psi(x)dx\right|_{x \to x + i0}\text{,}
\endmultline 
$$

which defines $C_{it}(x)$ as an analytic even functional on test functions $\psi(x) \in Z^{(1)}$: 
$$
\int \overline{C_{it}(x)}\psi(x)dx = {1\over{2\pi i}}\left(\int {{\psi(x + i0)}\over{t - (x + i0)}dx} + \int {{\psi(x - i0)}\over{t + (x - i0)}dx}\right)\text{.}
$$

That functional can be written as well as 
$$
\int \overline{C_{it}(x)}\psi(x)dx = {1\over{2\pi i}}\oint{{\psi(z)dz}\over{z - t}} + \left.{1\over{2\pi i}}\int \left({1\over{t - x}} + {1\over{t + x}}\right)\psi(x)dx\right|_{x \to x - i0}
$$
where the integration of the first summand is performed along the boundary of a rectangular strip that contains the real line. Thus 
$$
C_{it}(x) = \delta(x - t) + {i\over \pi} \cdot {t\over{t^2 - (x - i0)^2}} ,
$$
and, as well, 
$$
C_{it}(x) = {1\over 2}(\delta(x - t) + \delta(x + t)) + {i\over{2 \pi}}\left({t\over{t^2 - (x - i0)^2}} + {t\over{t^2 - (x + i0)^2}}\right). 
\tag 9
$$
It is clear that the other functional in (5) is $\overline{C}_{it}(x) = C_{-it}(x)$.
\medskip

{\it Remark.} The   functional $C_{it}(x)$ we have constructed can be also viewed as a functional on the space of bump test functions $\varphi(x) \in K^{(1)}$. Then 
$$
C_{it}(x) = {1\over 2}(\delta(x - t) + \delta(x + t)) + {i\over{\pi}} \cdot {t\over{t^2 - x^2}} . 
\tag 10
$$
Here the functional 
$$
{t\over{t^2 - x^2}} = {1\over 2}\left({1\over{t - x}} + {1\over{t + x}}\right)
$$
regularizes, and the integral $\int{{t\varphi(x)}\over{t^2 - x^2}}dx$ is understood as Cauchy's principal value, see \cite{2} and below. By abuse of notation, we denote the regularized functional by $C_{it}(x)$ as well. A reason for this is that the Fourier transform $\tilde{C}_{it}(p)$ of that functional, as seen from Parseval's identity 
$$
\int \overline{C_{it}(x)}\varphi(x)dx = {1\over{2\pi}}\int \left(\int \overline{C_{it}(x)}\exp(-ipx)dx\right)\psi(p)dp ,
\tag 11
$$
where $\psi(p)\in Z^{(1)}$ and the integral is understood as the principal value, is an analytic functional $\tilde{C}_{it}(p) = \exp(it|p|)$ on $
Z^{(1)}$.
\medskip

Therefore we have proved the next 
\medskip

{\bf Theorem (I):} {\it The inverse Fourier transform of the functional $\exp(it|p|)$ on $Z^{(1)}$ is equal to the even functional $C_{it}(x) = {1\over 2}(\delta(x - t) + \delta(x + t)) + {i\over{ \pi}} \cdot {t\over{t^2 - x^2}}$ on the bump test functions space $\varphi(x) \in K^{(1)}$. We call that generalized function $C_{it}(x)$   one-dimensional quantum Cauchy functional.}
\medskip

Notice that $C_{it}(x)$ (see (10)) satisfies the Chapman-Kolmogorov equation 
$$
C_{i\tau}(x_\tau) * C_{i(t-\tau)} = C_{it}(x_t)
$$
(here $*$ is the convolution of generalized functions), the existence of the convolution follows from the structure of the Fourier image $\tilde{C}_{it}(p)$ as a functional on $Z^{(1)}$.

It is clear that the infinitesimal operators (generators) $J_C(x)$ and $J_{\overline{C}}(x)$ that correspond to $C_{it}(x)$ and $\overline{C}_{it}(x)$ are, respectively, $-{i\over{\pi x^2}}$ and ${i\over{\pi x^2}}$. Therefore the quantum Cauchy functionals $C_{it}(x)$ and $\overline{C}_{it}(x)$ satisfy the equations 
$$
{\pderby t}C_{it}(x) = -{i\over{\pi x^2}} * C_{it}(x) \text{,\quad} {\pderby t}\overline{C}_{it}(x) = {i\over{\pi x^2}} * \overline{C}_{it}(x)\text{,} 
\tag 12
$$
and they are fundamental solutions of these equations (see \cite{3}).

\head 3. The space Cauchy distribution in imaginary time and the massless Dirac particle \endhead

Consider now the momentum space retarded Green's function $\tilde{C}_{it}(p) = \exp(it\rho)$ in (5) which is the analytic continuation to imaginary time of the space Cauchy process transition probability $C_t(x)$ viewed in the momentum coordinates. It is natural to call this process the space Cauchy process in imaginary time. As in one-dimensional case, we first assume that the functional $\tilde{C}_{it}(p)$ is defined on the space of bump test functions $\varphi(p) \in K^{(3)}$ (see \cite{2}).
\medskip

We find $C_{it}(x)$ using Parseval's identity that recovers a functional $C_{it}(x)$ on $Z^{(3)}$ from its Fourier transform $\tilde{C}_{it}(p)$ which is a functional on $K^{(3)}$. Namely, we have 
$$
\int \overline{C_{it}(x)}\psi(x)dx = {1\over{(2\pi)^3}}\int \overline{\exp(it\rho)}\varphi(p)dp\text{,} 
\tag 13
$$
where $\varphi(p) \in K^{(3)}$ is a bump function and $\psi(x) \in Z^{(3)}$ is its Fourier transform which is an entire function of first order (see \cite{2}).
\medskip

Consider in more details the integral in the r.h.s.~of (13); since
$$
\varphi(p) = \int \exp(i(x,p))\psi(x)dx\text{,}
$$
one has 
$$
\int \overline{C_{it}(x)}\psi(x)dx = {1\over{(2\pi)^3}}\int \int \exp(-it\rho + i(x,p))dp\psi(x)dx.
$$

Rewriting the inner integral in spherical coordinates, we get 
$$
\int \overline{C_{it}(x)}\psi(x)dx = {1\over{(2\pi)^3}}\int \int\limits_0^{\infty} \int\limits_{S_1} \exp(i\rho(-t + (x,p_e)))\rho^2 dS_1(p_e)d\rho\psi(x)dx, 
\tag 14
$$
where $S_1$ is the unit sphere in 3-dimensional space, and $dS_1(p_e)$ its area element. Therefore the functional 
$$
{1\over{(2\pi)^3}}\int\limits_0^{\infty} \int\limits_{S_1} \exp(i\rho(t - (x,p_e)))\rho^2 dS_1(p_e)d\rho
$$
is the sought-for inverse Fourier transform of the functional $\exp(it\rho)$. We rewrite it using the fact that  $\psi(x) \in Z^{(3)}$ and an orthogonal change of variables $x = Ay$, $y_1 = (x,p_e)$ (so $y_2$, $y_3$ are in the plane defined by equation $y_1 = (x,p_e)$); clearly $\psi(x) = \xi(y) \in Z^{(3)}$. Set $\int \xi(y)dy_2 dy_3 = \Xi(y_1) \in Z^{(1)}$. Therefore 
$$
\multline
\int \int\limits_0^{\infty} \int\limits_{S_1} \exp(i\rho(-t + (x,p_e)))\rho^2 dS_1(p_e)d\rho\psi(x)dx = \\
= \int \int\limits_0^{\infty} \int\limits_{S_1} \exp(i\rho(-t + (x,p_e)))\rho^2 dS_1(p_e)d\rho\Xi(y_1)dy_1\text{.}
\endmultline
\tag 15
$$

Since $\exp(i\rho y_1)$ is an entire function of $y_1$, the shift of the integration path  into complex plane $y_1 \mapsto y_1 + i0$ does not change the integral; then the integral by $\rho$ converges absolutely and  uniformly, and one has 
$$
\multline
\int \int\limits_0^{\infty} \int\limits_{S_1} \exp(i\rho(-t + (x,p_e) + i0))\rho^2 dS_1(p_e)d\rho\Xi(y_1 + i0)dy_1 = \\
= \left.{2\over i} \int \int\limits_{S_1}  {{dS_1(p_e)}\over{(-t + (x,p_e))^3}}\Xi(y_1)dy_1\right|_{y_1 \to y_1 + i0} = \\
= \left.-i \int {\psderby{(x,p_e)}}\int\limits_{S_1} {{dS_1(p_e)}\over{-t + (x,p_e)}}\Xi(y_1)dy_1\right|_{y_1 \to y_1 +i0} = \\
= -i{\psderby t}\int\limits_{S_1} \int {{\Xi(y_1 + i0 )dy_1}\over{-t + (y_1 + i0)}}dS_1(p_e)\text{.} 
\endmultline
\tag 16
$$

Therefore

$$
\multline
{\psderby t}\int\limits_{S_1} \int {{\Xi(y_1 + i0)dy_1}\over{-t + (y_1 + i0)}}dS_1(p_e) = \\
-{1\over 2}{\psderby t} \int\limits_{S_1} \oint {{\Xi(z)dz}\over{-t + z}}dS_1(p_e) + \left.\int\limits_{S_1} \int {{\xi(y)dy}\over{(-t + (Ay,p_e))^3}}dS_1(p_e)\right|_{y_1 \to y_1 + i0} + \\
+ \left.\int\limits_{S_1} \int {{\xi(y)dy}\over{(-t + (Ay,p_e))^3}}dS_1(p_e)\right|_{y_1 \to y_1 - i0}\text{,}
\endmultline
$$

where the contour in the first integral goes around the real axis. Consider the first summand; by Cauchy's theorem one has 
$$
\multline
{\psderby t}\int\limits_{S_1} \oint {{\Xi(z)dz}\over{z - t}}dS_1(p_e) = 2\pi i \int\limits_{S_1} \Xi^{(2)}(t)dS_1(p_e) = \\
 = 2\pi i \int\limits_{S_1} \int \delta^{(2)}(-t + (x,p_e))\Xi((x,p_e))d(x,p_e)dS_1(p_e)\text{.}
\endmultline
$$
Use the fact that in 3-space one has ``flat waves decomposition of the $\delta$-function"
$$
\delta(x) = {{-1}\over{2(2\pi)^2}}\int\limits_{S_1} \delta^{(2)}((x,p_e))dS_1(p_e)
$$
(see \cite{2}) which is a solution, in the generalized functions language, of the Radon problem of reconstruction of $\psi(0)$ from the integrals of $\psi$ along all planes $(x,p_e) = 0$. Thus 
$$
\multline
{{-1}\over{2(2\pi)^2}}\int \int\limits_{S_1} \delta^{(2)}(-t + (x,p_e))dS_1(p_e)\psi(x)dx = \\
 = {{\int\delta^{S_t}(x)\psi (x)dx}\over{4\pi t^2}} = \overline{\psi (x)}^{S_t} = \overline{\xi (y)}^{S_t} , 
\endmultline
\tag17
$$
where $\delta^{S_t}(x)$ is $\delta$-function of the sphere of radius $t$ and center at 0, and $\overline{\psi (x)}^{S_t}$ is the average of $\psi (t)$ over that sphere.
\medskip

From this, by (16) we see that $C_{it}(x)$ is an analytic functional on $Z^{(3)}$  
$$
\multline
\int \overline{C_{it}(x)}\psi (x)dx = \overline{\psi (y)}^{S_t} - \\
- {i\over{2\pi^2}}\left(\left.\int\limits_{S_1}\int {{\Xi (y_1)dy_1dS_1 (p_e )}\over{(-t+(Ay,p_e))^3}}\right|_{y_1 \to y_1 +i0} + \left.\int\limits_{S_1}\int {{\Xi (y_1)dy_1dS_1 (p_e )}\over{(-t+(Ay,p_e))^3}}\right|_{y_1 \to y_1 -i0}\right)\text{.} 
\endmultline
\tag 18
$$
It is important that this analytic functional can be viewed on bump functions $\varphi(x) \in K^{(3)}$ as well; then the integration over the unit sphere can be performed explicitly, and one has 
$$
\int \overline{C_{it}(x)}\varphi (x)dx = \overline{\varphi (x)}^{S_t} - {i\over{\pi^2}}\int {t\over{(t^2 - r^2 )^2}}\varphi(x)dx\text{.} 
\tag 19
$$

{\it Remark.} The functional (19) can be constructed, using (10) and theorem of inverse Radon transformation for measurable functions, see \cite{4}%Chapter 1. 
\medskip

In order to define that functional on the whole space $K^{(3)}$ we need to regularize the functional ${t\over{(t^2 - r^2 )^2}}$. One proceeds as follows. 
\medskip

Consider the integral $ \int {{t\varphi (x)dx}\over{(t^2 - r^2 )^2}}$ in spherical coordinates
$$
\int {{t\varphi (x)dx}\over{(t^2 - r^2 )^2}} = \int\limits_0^{\infty} {{t}\over{(t^2 - r^2 )^2}}\int\limits_{S_r} \varphi (x) dS_r (x)dr = 4\pi t \int\limits_0^{\infty} {{r^2}\over{(t^2 - r^2 )}}\overline{\varphi (x)}^{S_r}dr,
$$
where $\overline{\varphi (x)}^{S_r} = \Phi (r)$ is the average of $\varphi (x)$ along the sphere of radius $r$ with center at 0. Then $\Phi (r)\in K^{(1)}$ is an even bump function, see \cite{2}. Therefore the integral 
$$
\int\limits_0^{\infty} {{4tr^2}\over{(t^2 - r^2 )^2}}\Phi (r)dr = {1\over 2}\int\limits_{-\infty}^{\infty} {{4tr^2}\over{(t^2 - r^2 )^2}}\Phi (r)dr
$$
is a functional defined on the subspace of even bump functions in $K^{(1)}$.
\medskip

One has 
$$
{{4tr^2}\over{(t^2 - r^2 )^2}} = {{t}\over{(t - r )^2}}- {{1}\over{t - r}} + {{t}\over{(t + r )^2}}-{{1}\over{t + r}}.
$$
Notice also that the functional
$$
\int\limits_{-\infty}^{\infty}{{\Phi(r)}\over{(t-r)^2}}dr = -\int\limits_{-\infty}^{\infty}{{\Phi^{(1)}(r)}\over{t-r}}dr
$$
is defined on arbitrary odd bump functions, hence the functional
$$
-\int\limits_{-\infty}^{\infty}{{t\Phi^{(1)}(r)}\over{t-r}}dr - \int\limits_{-\infty}^{\infty}{{\Phi (r)}\over{t-r}}dr = \int {{\varphi (r)}\over{t-r}}dr
$$
is defined on arbitrary bump functions $\varphi (r)\in K^{(1)}$. So we have reduced the regularization of the functional in (19) to the already considered regularization in dimension one (10).
\medskip

As in 1-dimensional case, we keep to denote the quantum Cauchy functional by $C_{it}(x)$ and its momentum coordinate presentation by $\tilde{C}_{it}(p)$, understood now as   functionals on $K^{(3)}$ and $Z^{(3)}$.
\medskip

Thus we have proved

{\bf Theorem (II):} {\it The inverse Fourier transform of the functional $\exp(it\rho )$ on $Z^{(3)}$ is equal to the spherically symmetric functional
$$
C_{it}(x) = {{\delta^{S_t}}\over{4\pi t^2}}+ {i\over{\pi^2}} \cdot {t\over{(t^2 - r^2 )^2}}
$$
on bump functions $\varphi (x)\in K^{(3)}$. We call this generalized function $C_{it}(x)$  the space quantum Cauchy functional. }
\medskip

{\bf Corollary 1.} {\it The fundamental solution of the Dirac equation for massless particle has structure $D_t (x) = T(x)* D^F_t (x) * T(x)$, where (see (3), (4), (5)) 
$$
D^F_t (x) = 
\pmatrix
C_{it}(x) &0&0&0\\
0&C_{it}(x)&0&0\\
0&0&\bar{C}_{it}(x)&0\\
0&0&0&\bar{C}_{it}(x)
\endpmatrix
\tag 25
$$
does not lie in the Minkowski world. The solutions of the Dirac equations (3) and the Dirac equations in the Foldy-Wouthuysen coordinates $D^F_t(x)$ are isomorphic.}
\medskip

The shape of $\tilde{C}_{it}(p)$ implies that $C_{it}(x)$, viewed as a functional on the space of bump functions $K^{(3)}$ is a retarded Green's function hence satisfies the Chapman-Kolmogorov equation 
$$
C_{i\tau}(x_\tau ) * C_{i (t-\tau}(x_{t-\tau}) = C_{it}(x_t ),
$$
where $0\le\tau\le t$. Notice that the convolution is well defined since $\tilde{C}_{it}(p) = \exp(it\rho )$ is a functional on $Z^{(3)}$.
\medskip

Notice also that the generator of this retarded Green's function $C_{it}(x)$ equals ${i\over{\pi^2}}r^{-4}$. Thus the functional 
$C_{it}(x)$ is a fundamental solution of an integral equation (see \cite{3}) 
$$
{\pderby t}  C_{it}(x) = {i\over{\pi^2 r^4}} * C_{it}(x). 
\tag 20
$$

The results of this work are based on the study of those quantum Cauchy functionals $C_{it}(x)$ and $\bar{C}_{it}(x)$.

\head 4. Einstein's photons and the quantum Cauchy functional \endhead

Consider the equations for electromagnetic field  without sources 
$$
{\pderby t} E = \rot H, \quad \div E = 0 , \quad {\pderby t} H = - \rot E,\,\,\,  \div H = 0 ; 
\tag 21
$$
here $E_t(x)$ and $H_t(x)$ are, respectively, electric and magnetic fields or a photon field. We study solutions of these equations in Majorana coordinates (see \cite{1}) for
$$
M_t(x) = E_t(x)+ iH_t(x) \text{,\quad} \bar{M}_t(x) = E_t(x)- iH_t(x) \text{.}
$$

The  equations for them in the momentum coordinates $\tilde{M}_t (p)$, $\bar{\tilde{M}}_t (p)$ are
$$
\eqalign{
&\ i{\pderby t} \tilde{M}_t(p) = (S,p)\tilde{M}_t(p) \text{,\quad} (p,\tilde{M}_t(p)) = 0, \cr
&\ i{\pderby t} \overline{\tilde{M}}_t(p) = (S,p)\overline{\tilde{M}}_t(p) \text{,\quad} (p,\overline{\tilde{M}}_t(p)) = 0\text{.} \cr
}
\tag 22
$$

Here $(S,p) = \mathop\sum\limits_{j = 1}^{3} s^j p_j$ where 
$$
s^1 = 
\pmatrix
0&0&0 \\
0&0&-i \\
0 & i &0
\endpmatrix \text{,\quad} 
s^2 = 
\pmatrix
0&0&i \\
0&0&0 \\
-i & 0 &0
\endpmatrix \text{,\quad} 
s^3 = 
\pmatrix
0&-i&0 \\
i&0&0 \\
0 & 0 &0
\endpmatrix
$$
are the photon spin operators, so the operator 
$$
(S,p) = 
\pmatrix
0&-ip_3&ip_2 \\
ip_3&0&-ip_1 \\
-ip_2&ip_1&0
\endpmatrix
\tag 23
$$

is Hermitian.
\medskip

Consider the system of equations (22). First notice that the conditions in (22) are automatically satisfied since ${\pderby t} (p,\tilde{M}_t(p)) = 0$ as follows from (22). 
\medskip

The roots of the characteristic polynomial of the Hermitian matrix $(S,p)$ are $\pm \rho$ ($\rho = \sqrt {p_1^2 + p_2^2 + p_3^2}$) and $0$, so this matrix is degenerate. Therefore 
$$
(S,p) = \tilde{Q}^+(p)\tilde{h}^F(p)\tilde{Q}(p) \text{,}
$$
where 
$$
\tilde{h}^F (p) = 
\pmatrix
\rho &0&0 \\
0&-\rho&0\\
0&0&0
\endpmatrix \text{,}
\tag 24
$$

$\tilde{Q} (p)$ is a unitary operator that diagonalizes $(S,p)$, and $\tilde{Q}^+ (p)$ is the conjugate operator.

Therefore the fundamental solution of (24) viewed in the momentum coordinates is the direct product $\tilde{\text{M}}_t (p) = \tilde{M}_t (p)\times \overline{\tilde{M}}_t (p)$ of matrices
$$
\tilde{M}_t (p) = \tilde{Q}^+ (p)\tilde{M}^F_t (p) \tilde{Q} (p), \quad  \tilde{M}^F_t (p) = 
\pmatrix
\exp(-it\rho) &0&0\\
0&\exp(it\rho) &0 \\
0&0&1
\endpmatrix \text{,}
\tag 25  
$$
and $\overline{\tilde{M}}_t (p) = \tilde{M}_{-t} (p)$ which are analytic functionals on $Z^{(3)}$. Therefore M$_t (x)$, as the position coordinate presentation of the generalized function $\tilde{\text{M}}_t (p)$, is a functional on $\varphi (x)\in K^{(3)}$, and we have 
\medskip

{\bf Corollary 2 of theorem (II).} {\it The fundamental solution of the Maxwell equation (3). viewed as a functional on $K^{(3)}$, has structure}
$$
M_t (x) = Q^+ (x) * M^F_t (x) * Q(x)\times Q^+ (x) * \overline{M}^F_t (x) * Q(x) \text{,}
$$
{\it where functional 
$$
M^F_t (x) = 
\pmatrix 
\bar{C}_{it}(x)&o&0\\
0&C_{it}(x)&0\\
0&0&\delta (x)
\endpmatrix  
$$
evidently does not lie in the Minkowski world. Yet the solutions of the Maxwell equation (3) and the Maxwell equation in the Foldy-Wouthuysen coordinates are unitary equivalent.}

\head 5. A modified one-dimensional Cauchy functional \endhead

Consider the Dirac equation for a mass $m$ particle (see \cite{8})
$$
\gamma^0 {\pderby t}   + (\gamma, \nabla ) - im = 0. 
\tag 26
$$
Its fundamental solution in the momentum coordinates is 
$$
\tilde{D}^m_t (p) = \exp(it (\gamma^0 (\gamma,p)+m\gamma^0)) . 
$$

Since the above inner bracket is an Hermitian matrix, it can be diaonalized by a unitary (and Hermitian) transform $\tilde{T}^m (p)$
$$
\tilde{T}^m(p) = \tilde{T}^{m^{-1}}(p) = \gamma^0{{{(\gamma,p) + I(m + \sqrt{m^2 + \rho^2})}}\over{\sqrt{2\sqrt{m^2 + \rho^2}(m + \sqrt{ m^2 +\rho^2})}}}, 
\tag 27
$$
hence  
$$
\tilde{D}^m_t (p) = \tilde{T}^m (p)\exp(it \gamma^0 \sqrt{m^2 +\rho^2}  ) \tilde{T}^m (p). 
\tag 28
$$

Here $\gamma^0 \sqrt{m^2 +\rho^2} $ is the  momentum coordinates Foldy-Wouthuysen presentation  for the energy of the Dirac electron (see \cite{13}), and ${\tilde{D}^{m^F}_t}(p) = \exp(it\gamma^0 \sqrt{m^2 + \rho^2})$ is the  momentum coordinates Foldy-Wouthuysen presentation for the fundamental solution of the Dirac electron equation, see \cite{13}.
\medskip

We will call 
$$
\tilde{C}^m_{it}(p) = \exp(it\sqrt{m^2 + \rho^2}) 
\tag 29
$$
{\it the momentum coordinates presentation of the modified quantum Cauchy functional}. It is clear that we need to know the position coordinate presentation of that functional in order to construct $D^m_t (x)$.
\medskip

Our task now is to compute the inverse Fourier transform of ${\tilde{D}^{m^F}_t} (p)$ in one-dimensional case; to do this, we compute the inverse  Fourier transform of the functional  $\tilde{C}^m_{it}  (p)$ as a generalized function on $Z^{(1)}$ by a method we used to solve a similar problem above. Namely, we find $C^m_{it}  (x)$ from Parseval's identity 
$$
\int \overline{C^m_{it}  (x)}\psi (x)dx = {1\over{2\pi}} \int\overline{\tilde{C}^m_{it}  (p)}\varphi (p)dp, 
\tag 30
$$
where $\psi (x) \i Z^{(1)}$ and $\varphi (p)\in K^{(1)}$. We compute  the right integral by deforming the path of integration to make it absolutely convergent 
$$
\gather
\int \exp(-it\sqrt{m^2 + p^2})\left(\int\exp(ixp)\psi(x)dx\right)dp = \\
 = \left.\int\int\limits_{-\infty}^0 \overline{\tilde{C}^m_{it}(p)}\exp(ixp)dp\psi(x)dx\right|_{x\to x-i0} + \left.\int\int\limits_0^{\infty} \overline{\tilde{C}^m_{it}(p)}\exp(ixp)dp\psi(x)dx\right|_{x\to x+i0} .
\endgather
$$

Since $\tilde{C}^m_{it}  (p)$ is even with respect to $p$ one has 
$$
\multline
\int\left(\int\exp(-it\sqrt{m^2 + p^2})\exp(ixp)dp\right)\psi (x)dx = \\
 = \left.\int\int\limits_0^{\infty} \overline{\tilde{C}^m_{it}  (p)}e^{-ixp}dp \psi (x)dx\right|_{x\to x-i0}+ \left.\int\int\limits_0^{\infty} \overline{\tilde{C}^m_{it}  (p)}e^{ixp}dp \psi (x)dx\right|_{x\to x+i0} = \\
 = \left.\int\int\limits_0^{\infty} \overline{\tilde{C}^m_{it}  (p)}\cos (xp)dp \psi (x)dx\right|_{x\to x-i0}+ \left.\int\int\limits_0^{\infty} \overline{\tilde{C}^m_{it}  (p)}\cos (xp)dp \psi (x)dx\right|_{x\to x+i0},  
\endmultline
\tag 31
$$
since the integrals with sine vanish by the Cauchy theorem.
\medskip

One has (see \cite{12} formula 3.914) 
$$
\int\limits_0^\infty\exp(-t\sqrt{p^2 +m^2})\cos (px)dp = {{tm}\over{\sqrt{t^2 +x^2}}} K_1 (m\sqrt{t^2 +x^2}),
\tag 32
$$
where $K_1(z)$ is the Macdonald function (see \cite{10}, 3.7, formula 6); here $t$ and $x$ are real and $t>0$.
\medskip

We view this  as an equality  of analytic functionals 
$$
\int \int\limits_0^\infty\exp(-t\sqrt{p^2 +m^2})\cos (px)dp\psi (x) dx = \int {{tm}\over{\sqrt{t^2 +x^2}}} K_1 (m\sqrt{t^2 +x^2})\psi (x) dx 
$$
where $\psi (x)\in Z^{(1)}$. Since cosine is an even function, we can move the integration path, so one has 
$$
\multline
\left.\int \int\limits_0^\infty\exp(-t\sqrt{p^2 + m^2})\cos(px)dp\psi(x)dx\right|_{x\to x\pm i0} = \\
= \left.\int {{tm}\over{\sqrt{t^2 +x^2}}}K_1(m\sqrt{t^2 +x^2})\psi(x)dx\right|_{x\to x\pm i0}.
\endmultline
$$
The latter equality  can be continued analytically to $t$ in the imaginary axis:   
$$
\multline
\left.\int \int\limits_0^\infty \exp(-it\sqrt{p^2 + m^2})\cos(px)dp\psi(x)dx\right|_{x\to x\pm i0} = \\
= \left.\int {{tm}\over{\sqrt{t^2 -x^2}}}K_1(im\sqrt{t^2 -x^2})\psi(x)dx\right|_{x\to x\pm i0}. 
\endmultline
$$
Thus (31) becomes 
$$
\gather
\int\int\overline{\tilde{C}^m_{it}(p)}\exp(ixp)dp \psi (x)dx = \\
 = \left.\int {{tmK_1(im\sqrt{t^2 -x^2})}\over{\sqrt{t^2 -x^2}}}\psi (x) dx\right|_{x\to x- i0}\,\,+ \left.\int {{tmK_1(im\sqrt{t^2 -x^2})}\over{\sqrt{t^2 -x^2}}}\psi (x) dx\right|_{x\to x+ i0}, 
\endgather
$$
and Parseval's identity (30) yields  
$$
\gathered
\int\overline{C^m_{it}(x)}\psi (x)dx = \\
= {1\over{2\pi}}\oint {{tmK_1(im\sqrt{t^2 -z^2})}\over{\sqrt{t^2 -z^2}}}\psi(z)dz + \left.{1\over\pi} \int {{tmK_1(im\sqrt{t^2 -x^2})}\over{\sqrt{t^2 -x^2}}}\psi(x)dx\right|_{x\to x+ i0}\text{,} 
\endgathered
\tag 33
$$
where the first integral is taken along the boundary of an infinite rectangular strip that contains the real axis.
\medskip

Since 
$$
\left.K_1 (z)\right|_{z\to 0} \simeq z^{-1} 
\tag 34
$$
(see \cite{10}, 3.7, formulas (6), (2)), the function under the contour integral sign in (33) has simple poles at points $t$ and $-t$. So, by Cauchy's theorem, one has 
\medskip

$$
\int\overline{C^m_{it}(x)}\psi (x)dx = \left.{1\over{2 }}(\psi (-t) +\psi (t))+{1\over\pi} \int {{tmK_1(im\sqrt{t^2 -x^2})}\over{\sqrt{t^2 -x^2}}}\psi (x) dx\right|_{x\to x+ i0}.
\tag 35 
$$

The functional  $ C^m_{it}  (x)$, just as $ C_{it}  (x)$ above, can be seen to be defined on all the bump functions $\varphi (x)\in K^{(1)}$; then the integral in the r.h.s.~should be regularized since, by (34), the function under the integral has  simple poles on the real line at $t$ and $-t$.
\medskip

This can be done as follows. Set (see (10) and (35)) 
$$
{{tm}\over{\pi\sqrt{t^2 -x^2}}} K_1 (im\sqrt{t^2 -x^2})\left({{-i}\over {\pi}}\cdot {{t}\over{ {t^2 -x^2}}}\right)^{-1} = B(t,x); 
\tag 36
$$
then $B(t,x)$ and $B(t,x)^{-1}$ are infinitely differentiable functions that have no zeros in any finite domain with $t\ge 0$. And (34) implies that $B(t,\pm t ) = 1.$
\medskip

Consider the functional $C_{it}(x)B(t,x)$; one has 
$$
\int\overline{C_{it}(x)B(t,x)}\varphi (x) dx = {1\over 2} (\varphi (t)+ \varphi (-t))+ {{i}\over {\pi}}\int {{tmK_1 (im\sqrt{t^2 -x^2})}\over{\sqrt{t^2 - x^2 }}}\varphi (x) dx.
$$
Thus $\int \overline{C_{it}(x)B(t,x)}\varphi (x) dx = \int \overline{C^m_{it}(x)}\varphi (x) dx$ (see (36)), where $\varphi (x)\in K^{(1)}$, which means that 
$$
C_{it}(x)B(t,x) = C^m_{it}(x). 
\tag 37
$$
One also has $C^m_{it}(x) B^{-1}(t,x) = C_{it}(x).$
\medskip

Therefore, by (37), the regularization of the functional 
$$
\int \overline{C^m_{it}(x)}\varphi (x)dx = {1\over 2} (\varphi (t)+ \varphi (-t)) - {{i}\over {2\pi}}\int\left({1\over{t-x}}+{1\over{t+x}}\right)B(t,x)\varphi (x) dx
$$
is reduced to the interpretation of the integrals in the r.h.s.~as the integrals in the sense of Cauchy's principal value.
\medskip

Also, using (34), we get a relation between the functionals $C^m_{it}(x)$ and $C_{it}(x)$:  
$$
\lim_{m\to 0} C^m_{it}(x) = C_{it}(x). 
\tag 38
$$
Note that to the modified quantum Cauchy functional   $C^m_{it}(x)$ there corresponds a generator $J_{C^m} (x) = {{i}\over {\pi}}\cdot {{mK_1 (mx)}\over x}$, so the next equation (see \cite{3}) is satisfied 
$$
{\pderby t} C^m_{it}(x) = {{i}\over {\pi}}\cdot {{mK_1 (mx)}\over x} * C^m_{it}(x), 
\tag 39
$$
and $C^m_{it}(x)$ is its fundamental solution. 
\medskip

Therefore $C^m_{it}(x)$ satisfies the Chapman-Kolmogorov equation 
$$
C^m_{i\tau}(x_{\tau})* C^m_{i(t-\tau )}(x_{t-\tau}) = C^m_{it}(x_t),
$$
where the convolution is well defined due to the structure of $\tilde{C}^m_{it}(p) $ as a functional on $Z^{(1)}$. The same is true for $\bar{C}^m_{it}(x)$.
\bigskip

Thus we have proved 
\medskip

{\bf Theorem (III):} {\it The inverse Fourier transform of the functional
$$
\exp(it \sqrt{m^2 + p^2})
$$
on $Z^{(1)}$ is equal to the even functional
$$
C^m_{it}(x) = {1\over 2} (\delta (t-x)+ \delta (t+x))+ {{1}\over {\pi}} \cdot {{-tmK_1 (im\sqrt{t^2 -x^2})}\over{\sqrt{t^2 - x^2 }}}
$$
on $K^{(1)}$. We call this generalized function $C^m_{it}(x)$   the one-dimensional modified quantum Cauchy functional.}

\head 6. The  modified space quantum Cauchy  functional and the free Dirac electron \endhead

We construct now the inverse Fourier transform of the functional $\tilde{C}^m_{it}(p)$ as a generalized function on $Z^{(3)}$ using the same method as was used to solve the similar problem for the quantum Cauchy functional. Namely, we find $C^m_{it}(x)$ from Parseval's identity 
$$
\int \overline{C^m_{it}(x)}\psi (x)dx = {1\over{2\pi )^3}} \int\overline{\exp(it \sqrt{m^2 +\rho^2})}\varphi (p)dp, 
$$

where $\varphi (p)\in K^{(3)}$ and $\psi (x)\in Z^{(3)}$. Since $\varphi (p) = \int\exp(i(x,p))\psi (x)dx$, we see, rewriting the integral in spherical coordinates, that 
$$
\multline
\int\int\limits_0^{\infty} \overline{\exp(it \sqrt{m^2 +\rho^2})}\rho^2 \int\limits_{S_1}\exp(i\rho (x,p_e ))dS_1 (p_e )d\rho \psi (x)dx = \\
 = \int\int\limits_0^{\infty} \overline{\exp(it \sqrt{m^2 +\rho^2})}\rho^2 \int\limits_{S_1} \cos (\rho (x,p_e ))dS_1 (p_e )d\rho \psi (x)dx = \\
 = -\int\int\limits_0^{\infty} \overline{\exp(it \sqrt{m^2 +\rho^2})} \int\limits_{S_1}\left({\psderby{(x,p_e)}} \cos (\rho (x,p_e ))\right)dS_1 (p_e )d\rho \psi (x)dx 
\endmultline
\tag 40
$$
(here we use $\int\limits_{S_1} \sin (\rho (x,p_e ))dS_1 (p_e ) = 0$).
\medskip

As in (10), we perform the orthogonal change of variables $x = Ay$, and set $y_1 = (x,p_e )$, $\xi (y ) = \psi (x)$, and $\Xi (y_1 ) = \int \xi (u) dy_2 dy_3 \in Z^{(1)}$.
\medskip

Since $cos (\rho y_1 )$ is an even function, we can shift the integration path from the real axis $y_1$ in (40) to the complex domain $y_1 \to y_1 +i0$, so  
$$
\multline
\left.\int\limits_0^{\infty} \overline{\exp(it \sqrt{m^2 +\rho^2})}  \int\limits_{S_1}\int \left({\psderby{y_1}} \cos (\rho y_1)\right)\Xi(y_1 ) dy_1 dS_1 (p_e )d\rho\right|_{y_1 \to y_1 +i0} = \\
 = \left.\int\limits_{S_1}\int  \int\limits_0^{\infty} 
\overline{\exp(it \sqrt{m^2 +\rho^2})}   \cos (\rho y_1 )d\rho \Xi^{(2)} (y_1 ) dy_1 dS_1 (p_e ) \right|_{y_1 \to y_1 +i0}.
\endmultline
\tag 41
$$

Interpreting (32) as an equality of analytic functionals, we get 
$$
\multline
\left.\int\int\limits_0^\infty\exp(-t\sqrt{m^2 + \rho^2})\cos (\rho y_1 )d\rho \Xi^{(2)}(y_1 ) dy_1\right|_{y_1 \to y_1 +i0} = \\
 = \left.\int{{tm}\over{\sqrt{t^2 +y_1^2}}} K_1 (m\sqrt{t^2 +y_1^2})\Xi^{(2)}(y_1 ) dy_1\right|_{y_1 \to y_1 +i0}.
\endmultline
\tag 42
$$

As a result of analytic continuation by $y_1$, we can analytically continue the above equality to $t$ on the imaginary line:  
$$
\multline
\left.\int\int\limits_0^\infty\exp(-it\sqrt{m^2 + \rho^2})\cos (\rho y_1 )d\rho \Xi^{(2)}(y_1 ) dy_1\right|_{y_1 \to y_1 +i0} = \\
 = \left.\int{{tm}\over{\sqrt{t^2 - y_1^2}}} K_1 (im\sqrt{t^2 -y_1^2})\Xi^{(2)}(y_1 ) dy_1\right|_{y_1 \to y_1 +i0}.
\endmultline
\tag 43
$$

Therefore (40) implies that 
$$
\multline
\int\int\limits_0^{\infty}\exp\left(-it \sqrt{m^2 +\rho^2}\right)\rho^2 \int\limits_{S_1}\exp\left(i\rho (x,p_e )\right)dS_1 (p_e )d\rho \psi (x)dx = \\
 = \left.\int\limits_{S_1} \int {{tm}\over{\sqrt{t^2 - y_1^2}}} K_1\left(im\sqrt{t^2 -y_1^2}\right)\Xi^{(2)} (y_1 ) dy_1 dS_1 (p_e )\right|_{y_1 \to y_1 +i0}.
\endmultline 
\tag 44
$$

Since the function we integrate in (44) is analytic and $\Xi^{(2)}\in Z^{(1)}$, one has 
$$
\multline
\left.\int\limits_{S_1} \int {{tm}\over{\sqrt{t^2 - y_1^2}}} K_1\left(im\sqrt{t^2 -y_1^2}\right)\Xi^{(2)} (y_1 ) dy_1 dS_1 (p_e )\right|_{y_1 \to y_1 +i0} = \\
 = -{1\over 2}\int\limits_{S_1} \oint {{tm}\over{\sqrt{t^2 - z^2}}} K_1\left(im\sqrt{t^2 -z^2}\right)\Xi^{(2)} (z ) dz dS_1 (p_e )+ \\
+\left.{1\over 2}\int\limits_{S_1} \int {{tm}\over{\sqrt{t^2 - y_1^2}}} K_1\left(im\sqrt{t^2 -y_1^2}\right)\Xi^{(2)} (y_1 ) dy_1 dS_1 (p_e )\right|_{y_1 \to y_1 +i0}+ \\
+\left.{1\over 2}\int\limits_{S_1} \int {{tm}\over{\sqrt{t^2 - y_1^2}}} K_1\left(im\sqrt{t^2 -y_1^2}\right)\Xi^{(2)} (y_1 ) dy_1 dS_1 (p_e )\right|_{y_1 \to y_1 -i0}, 
\endmultline
\tag 45
$$

where the integration in the first summand is along a contour around the real axis. Consider that summand; then (34) implies that  the function we integrate has poles at $t$ and $-t$. By the Cauchy theorem 
$$
\oint {{tm}\over{\sqrt{t^2 - z^2}}} K_1\left(im\sqrt{t^2 -z^2}\right)\Xi^{(2)} (z ) dz = \pi \left(\Xi^{(2)} (t)+\Xi^{(2)} (-t)\right). 
\tag 46
$$

This implies 
$$
\multline
-{1\over 2}\int\limits_{S_1} \oint {{tm}\over{\sqrt{t^2 - z^2}}} K_1\left(im\sqrt{t^2 -z^2}\right)\Xi^{(2)} (z ) dz dS_1 (p_e ) = \\
 = -{\pi\over 2}\int\limits_{S_1} \int \delta^{(2)}\left(t+(x,p_e )\right) \left(\Xi \left((x,p_e )\right)+\Xi \left(-(x,p_e )\right)\right)d (x,p_e )dS_1 (p_e ) = \\
 = -\pi \int \int\limits_{S_1}\delta^{(2)}\left(t+(x,p_e )\right)   dS_1 (p_e )  \Xi \left((x,p_e )\right) d (x,p_e )
\endmultline
\tag 47
$$

Therefore the "decomposition of $\delta$-function by flat waves" we have already used (see \cite{2}) shows that $C^m_{it}(x)$ is an analytic functional on $Z^{(3)}$:
$$
\multline
\int\overline{C^m_{it}(x)}\psi (x)dx = \overline{\psi(x)}^{S_t} + \\
+ \left.{1\over{2(2\pi)^3}}\int\limits_{S_1} \int \left({\psderby{y_1}} {{tm K_1 \left(im\sqrt{t^2 -y_1^2}\right) }\over{\sqrt{t^2 - y_1^2}}}\right)\xi(y)dydS_1(p_e )\right|_{y_1 \to y_1 +i0}\,+ \\
+ \left.{1\over{2(2\pi)^3}}\int\limits_{S_1} \int  \left({\psderby{y_1}} {{tm K_1\left(im\sqrt{t^2 -y_1^2}\right) }\over{\sqrt{t^2 - y_1^2}}}\right)\xi(y)dydS_1 (p_e )\right|_{y_1 \to y_1 -i0} . 
\endmultline
\tag 48
$$
If we consider that functional on $K^{(3)}$, then integrals along $S_1$ in the last two summands can be computed, and one has 
$$
\multline
\int\limits_{S_1}{\psderby{(x,p_e)}} {{tm K_1\left(im\sqrt{t^2 -(x,p_e )^2}\right) }\over{\sqrt{t^2 - (x,p_e )^2}}} dS_1 (p_e ) = \\
= 2\pi r^{-1} \int\limits_{-r}^r {\psderby{y_1}} {{tmK_1\left(im\sqrt{t^2 -y_1^2}\right)}\over{\sqrt{t^2 - y_1^2}}}dy_1 = \\
= -8\pi {\pderby{(t^2 - r^2)}} {{tm K_1\left(im\sqrt{t^2 -r^2}\right)}\over{\sqrt{t^2 - r^2}}}, 
\endmultline
\tag 49
$$
where $r = |y|$.
\medskip

Notice that on bump functions $\varphi (x)\in K^{(3)}$ that functional 
$$
\int\overline{C^m_{it}(x)}\varphi (x)dx = \overline{\varphi (x)}^{S_t}+ {i\over{\pi^2}}\int {\pderby{(t^2 - r^2)}} {{-tm K_1\left(im\sqrt{t^2 -r^2}\right) }\over{\sqrt{t^2 - r^2}}}\varphi (x)dx 
\tag 50
$$

requires a regularization since the function we integrate has singularity on the sphere of radius $t$. To find it, we first rewrite that functional.
\medskip

{\it Remark.} And as in case space Cauchy functional (19) the functional (50) can be constructed by the theorem of inverse Radon transformation (35) for measurable functions, see \cite{4}.
\medskip

We use the regularized solution $C_{it}(x)$ in the space case (see (19)) that was deduced from the one-dimensional situation (12).
\medskip

Consider the function 
$$
{1\over{\pi^2}}\left( {\pderby{(t^2 - r^2)}} {{tm K_1\left(im\sqrt{t^2 -r^2}\right) }\over{\sqrt{t^2 - r^2}}}\right)\left( {i\over{\pi^2}} {t\over{( t^2 -r^2)^2 }}\right)^{-1} =
\tag 51
$$
$$ = iml^4 {\pderby{(l^2)}} {{K_1(iml)}\over l} = B(t,r) ,
$$
where $\sqrt{t^2 - r^2} = l$. That function is infinitely differentiable with respect to $r$, and $\lim\limits_{r\to t} B(t,r) = 1$ by (34). Both $B(t,r)$ and $B(t,r)^{-1}$ do not vanish.
\medskip

Consider $ C_{it}(x)B(t,r)$ (see (19)) as a functional on $K^{(3)}$; then 
$$
\int\overline{C_{it}(x)B(t,r)}\varphi (x)dx = \int\overline{C^m_{it}(x)} \varphi (x)dx 
\tag 52
$$
for every $\varphi (x)\in K^{(3)}$. Hence 
$$
C_{it}(x)B(t,r) = C^m_{it}(x) .
\tag 53
$$
Therefore 
$$
\int  \overline{C^m_{it}(x)} \varphi (x)dx = \overline{\varphi (x)}^{S_t} - {i\over{\pi^2}}\int{1\over{( t^2 -r^2)^2 }}B(t,r)\varphi (x)dx , 
\tag 54
$$
so the asked for regularization of the integral in (54) is reduced to the already performed regularization of the functional $C_{it}(x)$ (see (19)).
\medskip

We notice also that there is a natural relation 
$$
\lim_{m\to 0} C^m_{it}(x) = C_{it}(x).
$$

It is clear that the functional $ C^m_{it}(x)$ is the fundamental solution of an integral equation 
$$
{\pderby t}C^m_{it}(x) = {{im}\over {2\pi^2}}{1\over r} {\pderby r} {{K_1 (mr)}\over r} *       C^m_{it}(x) , 
\tag 55
$$
that satisfies the Chapman-Kolmogorov equation 
$$
C^m_{i\tau}(x_\tau ) * C^m_{i(t-\tau )}(x_{t-\tau}) = C^m_{it}(x_t), 
\tag 56
$$
the convolution is well defined due to the structure of $\tilde{C}^m_{it}(p) = \exp(it\sqrt{m^2 +\rho^2 })$. It is easy to see that $\bar{C}^m_{it}(x) = C^m_{-it}(x)$ has similar properties.
\bigskip

Thus we have deduced 
\medskip

{\bf Theorem (IV):} {\it The inverse Fourier transform of the functional
$$
\exp(it\sqrt{m^2 +\rho^2 })
$$
on $Z^{(3)}$ is the spherically symmetric functional
$$
C^m_{it} (x) = {{\delta^{S_t}}\over{4\pi t^2}} + {1\over{\pi^2}} {\pderby{(t^2 - r^2)}} {{tm K_1 (im\sqrt{t^2 -r^2}) }\over{\sqrt{t^2 - r^2}}}
$$
on $K^{(3)}$. We call that functional  the modified space quantum Cauchy functional.}
\medskip

{\bf Corollary.} {\it The fundamental solution of the Dirac equation (1) is a matrix-valued functional $D^m_t (x) = T^m (x)* D^{m^F}_t (x) *T^m (x)$ on $K^{(3)}$ (see (3), (4), (5)), where 
$$
D^{m^F}_t(x) = 
\pmatrix
C^m_{it} (x)&0&0&0\\
0&C^m_{it} (x)&0&0\\
0&0&\bar{C}^m_{it} (x)&0\\
0&0&0&\bar{C}^m_{it} (x)
\endpmatrix 
$$
clearly does not lie in Minkowski's world. Yet  the solutions of the Dirac equation (26) and the Dirac equation in the Foldy-Wouthuysen $D^{m^F}_t(x)$ coordinates are unitary equivalent.}
\bigskip

Therefore the evolution of Dirac's electron is reduced to its evolution in the Foldy-Wouthuysen coordinates (that preserves spin), after which the operations of the  left and right convolutions with $T^m (x)$ return the spin to the construction.

\head 7. On the correspondence principle for the Dirac electron \endhead

Consider now the problem of construction of the quasi-classical solution to the Dirac electron equations in more details (cf.~\cite{11}).
\medskip

Recall that the fundamental solution of the Dirac electron equation in the momentum coordinates is $\tilde{D}^m_t (p) = \tilde{T}^m (p)\tilde{D}^{m^F}_t (p) \tilde{T}^m (p)$ where
$$
\tilde{D}^{m^F}_t (p) = \exp(it\gamma^0 \sqrt{m^2 + \rho^2})
$$
is the Foldy-Wouthuysen  solution, and $\tilde{T}^m (p)$ is as in (27). To return to physical coordinates we have to change $m$ by $\hbar^{-1}cm_0$, where $m_0$ is   the invariant mass of the electron, and $t$ by $ct$ (see \cite{13}); then our quasi-classical approximation is reduced to finding of the quasi-classical approximation $D^m_t (x)$ and $D_t^{m^F} (m)$ for $m\to\infty$; here, as before, $c = 1$.
\medskip

If  $\hbar \to 0$ then  $\tilde{T}^m (p)\to I$ hence $\tilde{D}^m_t (p)\to \tilde{D}^{m^F}_t (p)$, but, despite the vanishing of the operator responsible for the finiteness of the functional $D^m_t(x)$ (see \cite{8}), the quasi-classical limit for the   fundamental solution of the Dirac electron equation and its Foldy-Wouthuysen transform coincide. Let us show this.
\medskip

Notice that, due to the asymptotic of the Macdonald function
$$
\left.K_1 (z)\right|_{z\to\infty} \simeq \sqrt{\pi\over{2z}}\exp(-z)
$$
(see \cite{10} 7.2.3, formula (1)), yield the asymptotic of $\left.C^m_{it}(x)\right|_{\hbar \to 0}$:  as an analytic functional on $Z^{(3)}$ it is an entire function 
$$
\multline
\left.\int \overline{C^m_{it}}\psi(x)dx\right|_{\hbar \to 0} = \\
{{-t\sqrt{\pi/2}}\over{(2\pi)^3}} \left.\int \int\limits_{S_1} \left({\psderby{y_1}} {{\exp(i\hbar^{-1}m_0 l_t)}\over{{l_t}^{3/2}}}\right)dS_1(p_e)\Xi(y_1)dy_1\right|_{y_1 \to y_1 + i0} = \\
{{-t\sqrt{\pi/2}}\over{(2\pi)^2}} \left.\int \int\limits_0^\pi {\psderby{y_1}} {{\sin\theta \exp(i\hbar^{-1}m_0 l_t)}\over{{(t^2 - (r\cos\theta)^2)}^{3/2}}}d\theta \Xi(y_1)dy_1\right|_{y_1 \to y_1 + i0}\text{,}
\endmultline
$$
(here $l_t = \sqrt{t^2 - r^2}$ is the relativistic interval and $m_0 l_t$ is the free particle eikonal).
\medskip

Therefore, for $\hbar\to 0$ the poles at $t$, $-t$ disappear (hence the $\delta$-function in (50) disappears) and the integral over the domain $r\ge t$ (the complement to the light cone) disappears in two (of the four) diagonal elements of $D_t^{m^F}$, due to the exponential decrease of these in that domain.
\medskip

Therefore, the obtained functional can be considered as a functional on finite functions, that allow to calculate the integral by the unit sphere
$$
\multline
\left.\int \overline{C^m_{it}}\varphi(x)dx\right|_{\hbar \to 0} = \\
\left.{{t\sqrt{2\pi}}\over{2(2\pi)^3}} \int \left({\pderby{({l_t}^2)}} {{\exp(i\hbar^{-1}m_0 l_t)}\over{{l_t}^3}}\right)\varphi(x)dx\right|_{\hbar \to 0} = \\
\left.\int\exp({i\over{\hbar}}m_0 l_t )\phi(x)dx\right|_{\hbar \to 0}\text{,}
\endmultline
\tag 57
$$
where, easy to see that $\forall \phi(x) \in K^{(3)}$.
\bigskip 

So $C^m_{it}(x)|_{\hbar \to 0}$ becomes a finite functional that describes the classical free relativistic particle of an arbitrary finite mass, cf.~\cite{14}, \cite{8}, \cite{9}. 
\medskip 

We emphasize that the classical relativistic limit for the quantum relativistic particle of nonzero mass comes from the imaginary summand of Green's functional which has unbounded support.
\medskip

This implies, in particular, that it is impossible to interpret the quantization of a classical relativistic particle as consideration of arbitrary trajectories in the classical action integral, as it happens for the nonrelativistic theory (see \cite{9}, \cite{11}). Notice also that in quantum relativistic  case, as follows from (57), the dependence from the eikonal is exponential only in the quasi-classical approximation. This characteristic property of quantization of the classical relativistic particles should be
accounted for in the quantum theory.

\head Conclusion \endhead

The present work shows the special role of the quantum Cauchy functionals (which are the imaginary time Cauchy distributions) on the bump functions for understanding the solutions of the  relativistic quantum mechanics equations.
\medskip

These quantum Cauchy functionals appear naturally when one constructs solutions of the Dirac and Maxwell equations in the position coordinates in the Foldy-Wouthuysen form. Yet the physical facts, that correspond to these functionals, are beyond the casual classical Minkowski world.
\medskip

With the help of the quantum Cauchy functionals, we observe a fundamental relation between solutions of the Dirac and Maxwell equations and find unitary transformations that interchange bosons and fermions without leaving the classical Minkowski world, which may be related to  a possibility of construction of the F.~A.~Berezin  superinteraction theory (see \cite{6}). 
\medskip

We also find that these functionals can be effectively applied to the study of  the passage from  quantum relativistic problems to their classical relativistic versions.
\bigskip

The author thanks O.~G.~Smolyanov for stimulating support.

\end